\documentclass[floatfix,aps,twocolumn,showpacs,superscriptaddress,prl,10pt]{revtex4-1}
\usepackage{amsmath,amsfonts,amssymb,graphicx}
\usepackage{bm}

\usepackage[utf8]{inputenc}
\usepackage[T1]{fontenc}

\IfFileExists{newtxtext.sty}
   {\usepackage{newtxtext,newtxmath}}
   {\IfFileExists{stix.sty}
      {\usepackage{stix}}
      {\IfFileExists{mathptmx.sty}
      {\usepackage{mathptmx}}{} } }

\usepackage{hyperref} 
\usepackage{color}

\begin{document}
\title{Quantum Critical Behavior of One-Dimensional Soft Bosons in the Continuum}

\author{Stefano Rossotti}
\affiliation{Dipartimento di Fisica, Universit\`a degli Studi di Milano, via Celoria 16, I-20133 Milano, Italy}
\author{Martina Teruzzi}
\affiliation{International School for Advanced Studies (SISSA), Via Bonomea 265, Trieste, Italy}
\affiliation{Dipartimento di Fisica, Universit\`a degli Studi di Milano, via Celoria 16, I-20133 Milano, Italy}
\author{Davide Pini}
\affiliation{Dipartimento di Fisica, Universit\`a degli Studi di Milano, via Celoria 16, I-20133 Milano, Italy}
\author{Davide Emilio Galli}
\affiliation{Dipartimento di Fisica, Universit\`a degli Studi di Milano, via Celoria 16, I-20133 Milano, Italy}
\author{Gianluca Bertaina}
\email{gianluca.bertaina@unimi.it}
\affiliation{Dipartimento di Fisica, Universit\`a degli Studi di Milano, via Celoria 16, I-20133 Milano, Italy}

\begin{abstract}
We consider a zero-temperature one-dimensional system of bosons interacting via the soft-shoulder potential in the continuum, typical of dressed Rydberg gases. 
We employ quantum Monte Carlo simulations, which allow for the exact calculation of imaginary-time correlations, and a stochastic analytic continuation method, to extract the dynamical structure factor.
At finite densities, in the weakly-interacting homogeneous regime, a rotonic spectrum marks the tendency to clustering. With strong interactions, we indeed observe cluster liquid phases emerging, characterized by the spectrum of a composite harmonic chain. Luttinger theory has to be adapted by changing the reference lattice density field. In both the liquid and cluster liquid phases, we find convincing evidence of a secondary mode, which becomes gapless only at the transition. In that region, we also measure the central charge and observe its increase towards $c=3/2$, as recently evaluated in a related extended Bose-Hubbard model, and we note a fast reduction of the Luttinger parameter. For 2-particle clusters, we then interpret such observations in terms of the compresence of a Luttinger liquid and a critical transverse Ising model, related to the instability of the reference lattice density field towards coalescence of sites, typical of potentials which are flat at short distances. Even in the absence of a true lattice, we are able to evaluate the spatial correlation function of a suitable pseudo-spin operator, which manifests ferromagnetic order in the cluster liquid phase, exponential decay in the liquid phase, and algebraic order at criticality.
\end{abstract}

\maketitle

Quantum phase transitions (QPT) \cite{sachdev_quantum_2000} play an intriguing role in many-body systems, due to the possibility of unveiling new exotic phases. 
The progress in the manipulation of ultracold gases allows for the exploration of QPTs, by engineering well-controlled synthetic quantum many-body systems, confined for example by optical lattices \cite{simon_quantum_2011,zhang_observation_2012} or in quasi-one-dimensional geometries \cite{cazalilla_one_2011,olshanii_atomic_1998,haller_realization_2009,haller_pinning_2010}. Recently, Rydberg atoms \cite{low_experimental_2012} have emerged as a new route to QPTs \cite{weimer_quantum_2008,schauss_observation_2012}. These are atoms in highly-excited electronic states, with a very large electronic cloud.
In particular, theoretical \cite{henkel_threedimensional_2010,cinti_defectinduced_2014,saccani_excitation_2012,lauer_transportinduced_2012} and experimental \cite{jau_entangling_2016,zeiher_manybody_2016,zeiher_coherent_2017} efforts have focused on ensembles of \textit{dressed} Rydberg atoms, which are superpositions of the ground state and the above mentioned excited states, coupled via a Rabi process. Their effective interaction can be a soft-shoulder potential, with a flat repulsion up to a radius $R_c$ related to the highly excited state, and a repulsive van-der-Waals tail at large distances \cite{henkel_threedimensional_2010,pupillo_strongly_2010,cinti_defectinduced_2014,balewski_rydberg_2014,macri_ground_2014,plodzien_rydberg_2017}. Quite interestingly, this repulsive interaction belongs to the class that has been recognized to induce cluster formation at high density in classical statistical mechanics \cite{likos_criterion_2001,mladek_formation_2006}, thanks to the relative freedom of particles at short distances. This has opened a recent flourishing of research on quantum cluster phases: in high dimensions, coexisting cluster crystal and superfluid order have been predicted, yielding supersolid behavior \cite{henkel_threedimensional_2010,cinti_defectinduced_2014,saccani_excitation_2012,ancilotto_supersolid_2013}, while in one dimension (1D), cluster Luttinger liquids (CLL) have been proposed on a lattice \cite{mattioli_cluster_2013,dalmonte_cluster_2015}.

\begin{figure}[bp]
\centering
\includegraphics[width = 0.9\columnwidth]{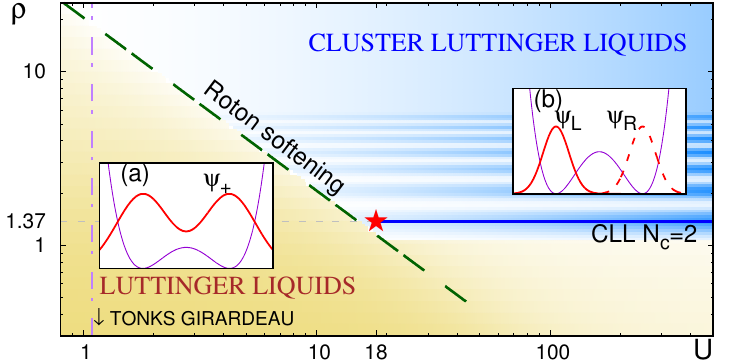}
\caption{Phase diagram (log-log scale). A star marks the critical point between the LL and CLL phases for densities commensurate to 2-particle clusters. The long-dashed line corresponds to the softening of the Bogoliubov roton. (a) Delocalized orbital in the effective double-well. (b) Localized left and right orbitals.}
\label{fig:phase} 
\end{figure}

In this Letter, we investigate a prototypical system of $N$ bosons in 1D at linear particle density $n$, governed by the following Hamiltonian in the continuum:
\begin{equation}\label{eq:hamiltonian}
 H = -\frac{\hbar^2}{2m}\sum_i^N \frac{\partial^2}{\partial x_i^2} + \sum_{i<j} \frac{V_0}{r_{ij}^6+R_c^6} 
\end{equation}
where $x_i$ are the particle coordinates, $r_{ij}=|x_i-x_j|$ the distances, $m$ is the mass and $V_0$ and $R_c$ are the strength and the radius of the soft-shoulder potential $V(r)$. 
If not otherwise specified, in the following we use units of $R_c$ for the length, $E_c=\hbar^2/m R_c^2$ for the energy, and $\hbar/R_c$ for the momenta (See Supplemental Material \footnote{See Supplemental Material at [URL will be inserted by publisher] for details on the potential, the experimental realization, the methods, the extraction of relevant observables.}). The zero-temperature phase diagram (Fig.~\ref{fig:phase}) thus depends on the following two dimensionless quantities: strength, $U=V_0/(E_c R_c^6)$, and density, $\rho=nR_c$. 
By evaluating relevant static and dynamical properties, we show that, while for small $U$ and moderate $\rho$ the system is a Luttinger liquid (LL) \cite{haldane_effective_1981}, although with strong correlation effects, for higher $U$ or $\rho$ a transition occurs towards CLL. In particular, we focus on the QPT to the dimer cluster liquid, which turns out to be of the 2D Ising universality class. A similar phenomenology has been recently studied in a 1D lattice system governed by the extended Bose-Hubbard hamiltonian \cite{mattioli_cluster_2013,dalmonte_cluster_2015}, while we observe it for the first time in the continuum, where we find that an effective spin hamiltonian emerges at the transition, \emph{even in absence of an underlying lattice}.

For generic coupling and density, this system falls into the LL universality class \cite{haldane_effective_1981}, characterized by a gapless bosonic mode at small momenta, with sound velocity $v$.
The low-energy and momentum sector of the Hilbert space is governed by the Hamiltonian $H_{LL}=(v/2\pi)\int dx [K_L (\nabla \theta)^2+(\nabla \phi)^2/K_L]$, where a large Luttinger parameter $K_L>1$ favors the fluctuations of the particle counting field $\phi(x)$, while small values of $K_L$ induce crystal-like behavior, by disordering the phase field $\theta(x)$. The central charge of the associated conformal field theory (CFT) is $c=1$ \cite{giamarchi_quantum_2003} and, for our Galilean invariant system, $v=\rho \pi/K_L$ \cite{haldane_effective_1981}.

In the dilute limit, the effects of the interaction are well described by the scattering length $a_{\text{1D}}$; in particular, for $U \sim 1.09$, we get $a_{\text{1D}} \sim 0$ \cite{[{}] [{. Recall that the definition of the scattering length in 1D implies that it is zero for infinite zero-range repulsion.}] teruzzi_microscopic_2017}, corresponding to the Tonks--Girardeau (TG) model \cite{girardeau_relationship_1960}. Conversely, at higher densities, the full shape of $V(r)$ is relevant. Its 1D Fourier transform $\tilde{V}(q)$ \cite{Note1} features a global minimum at $q_c \simeq 4.3$, at which $\tilde{V}(q_c)<0$, providing a typical length $b_c=2\pi/q_c \simeq 1.46$. It has been recognized, in the context of classical physics, that such density-independent distance favors clustering, even with a completely repulsive potential \cite{likos_criterion_2001,mladek_formation_2006}. 
Classically, one obtains a $T=0$ cluster crystal, which is destabilized in 1D by finite temperature, in favor of cluster-dominated liquid phases with different average occupation \cite{prestipino_cluster_2014,prestipino_probing_2015}. Quantum mechanics induces \emph{coherent} delocalization even at $T=0$, rendering the cluster phase a CLL and triggering a QPT towards a LL without cluster order (Fig.~\ref{fig:phase}).

\begin{figure}[pbt]
\centering
\includegraphics[width = \columnwidth]{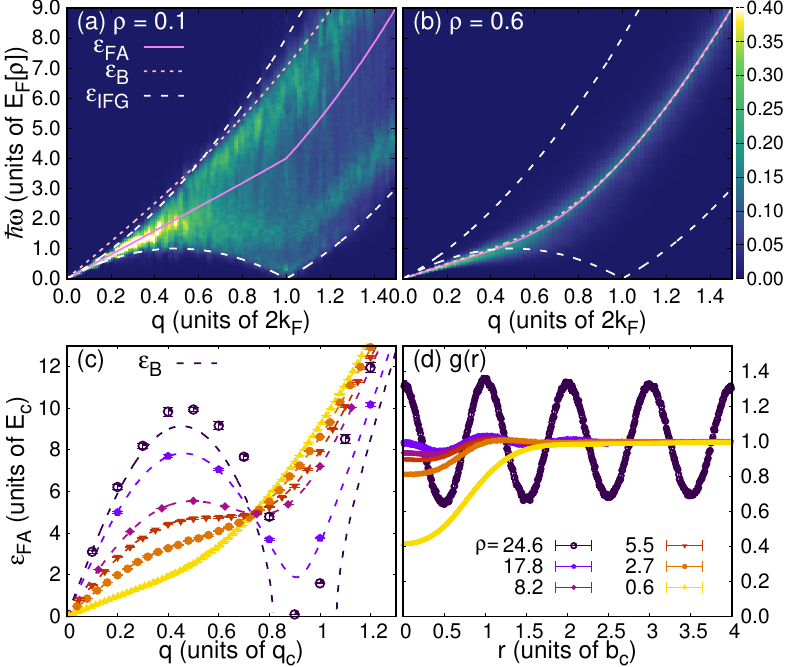}
\caption{Dynamical structure factor at $U=1.09$, as a function of momentum $q$ and energy transfer $\varepsilon=\hbar\omega$. Magnitude (color box) is in units of the inverse Fermi energy (values beyond scale are plotted in white). Feynman $\varepsilon_{FA}$ and Bogoliubov $\varepsilon_B$ approximations, and the ideal Fermi gas particle-hole boundaries $\varepsilon_{IFG}$, are plotted as a reference. Liquid phase at low (a) and intermediate $\rho$ (b). (c) Evolution of $\varepsilon_{FA}$, compared to $\varepsilon_{B}$. (d) Pair distribution function.}
\label{fig:spectraliquid} 
\end{figure}

To study the phase diagram in a non-perturbative way, we use the well-established path integral ground state (PIGS) quantum Monte Carlo method \cite{sarsa_path_2000,rossi_exact_2009}, which represents the ground state as the imaginary-time projection $\exp{(-\tau H)}|\Psi_{T}\rangle$ of a trial wavefunction. We simulate up to $N=200$ particles in a segment of length $L=N/\rho$, using periodic boundary conditions (PBC). The trial wavefunction is of the two-body Jastrow form: $\Psi_T(x_1\dots x_N) = \exp\left[-\frac{1}{2}\sum_{i<j} \left(u(r_{ij}) + \chi(r_{ij})\right)\right]$, where $\chi(r)$ accounts for long-wavelength phonons \cite{reatto_phonons_1967,teruzzi_microscopic_2017}, while $\exp\left[-u(r)/2\right]$ is the numerical solution of a two-body Schr\"odinger equation \cite{teruzzi_microscopic_2017}, with the effective potential $V_{eff}(r)=c_1 V(r) + c_2 \sum_l V(r-l b)$. To reduce projection times, it is crucial to use and optimize this mean-field potential, which accounts for the presence of nearby clusters \cite{Note1}.
We consider excitations associated to density fluctuations, which are commonly investigated via the dynamical structure factor $S(q,\omega) = \int dt \frac{e^{i \omega t}}{2\pi N} \langle e^{itH/\hbar}\rho_q e^{-itH/\hbar} \rho_{-q} \rangle$.
The PIGS algorithm evaluates the numerically exact imaginary-time intermediate scattering function, which yields $S(q,\omega)$ via analytic continuation, through the genetic inversion via falsification of theories algorithm \cite{vitali_initio_2010,bertaina_onedimensional_2016,motta_dynamical_2016,bertaina_statistical_2017,Note1}. 

We now proceed to discuss our results, first in the LL, then in the CLL regimes. Finally, we discuss the QPT in between the two liquids.

\paragraph{Liquid regime--} Here and in the following, $k_F=\pi\rho$ and $E_F=k_F^2/2$ are effective Fermi momentum and energy. We concentrate on interaction $U \simeq 1.09$, and increase the density (dot-dashed line in Fig.~\ref{fig:phase}). 
In the low-density regime $\rho\lesssim 0.1$ (Fig.~\ref{fig:spectraliquid}a), $S(q,\omega)$ is almost constant, at fixed $q$, in between the particle-hole boundaries $\varepsilon_{IFG}(q) = \left| k_F q \pm q^2/2\right|$, analogously to the TG gas, which can be mapped to an ideal Fermi gas (IFG). However, within our resolution, the spectral weight has started to gather, especially at the upper boundary, similarly to what happens in the Lieb-Liniger model with decreasing coupling parameter \cite{caux_dynamical_2006}.
In fact, already at $\rho=0.6$ (panel b),
the spectrum has evolved into a main mode.
As such, it is very well described by the single-peak Feynman approximation $\varepsilon_{FA}(q)=\varepsilon_0(q)/S(q)$, with $\varepsilon_0(q)=q^2/2$ the free-particle energy and $S(q)$ the static structure factor. 

By further increasing $\rho$, we simply monitor the evolution of $\varepsilon_{FA}(q)$ (Fig.~\ref{fig:spectraliquid}c), and notice that the main excitation becomes more structured, with a roton minimum moving towards $q=q_c$. A standard Bogoliubov analysis \cite{henkel_threedimensional_2010,macri_ground_2014} yields the dispersion $\varepsilon_B(q)=\sqrt{\varepsilon_0(q)\left[\varepsilon_0(q)+2 \rho \tilde{V}(q)\right]}$, which depends only on the combination $\alpha=\rho U$. In this approximation, it is clear that the emergence of the roton minimum is allowed by the momentum dependence of $\tilde{V}(q)$, which has a negative part \cite{santos_rotonmaxon_2003,odell_rotons_2003}. The roton softens at $\alpha=\alpha_c \simeq 20.65$. While the agreement between the single-mode $\varepsilon_{FA}(q)$ and $\varepsilon_B(q)$ approximations is very good for $0.6\lesssim\rho\lesssim 19$, such treatments are in general not valid anymore for $U \gtrsim \alpha_c/\rho$ (dashed line in Fig.~\ref{fig:phase}), where indeed our simulations show that clustering occurs.

On increasing $\rho$, the pair distribution function $g(r)$ at first gradually approaches 1 everywhere (Fig.~\ref{fig:spectraliquid}d), as in classical soft-core fluids in the absence of clustering \cite{lang_fluid_2000}. However, for very high $\rho$, large-amplitude slowly-decaying oscillations appear, with wavelength $b_c$. Again, this behavior is akin to that of classical systems, in the presence of clustering \cite{likos_criterion_2001}. 
A gaussian fit of the peaks indicates, on average, $N_c\simeq 36$ particles per cluster at $\rho=24.6$. In the quantum case, the oscillations of $g(r)$ eventually decay as in a cluster liquid, a behavior that we can easily see in the more relevant cluster phases at low $\rho$ and large $U$. In fact, a hamiltonian description of dressed Rydberg gases is questionable at high $\rho$, due to increased losses to other Rydberg levels in current experiments \cite{Note1}.

\begin{figure}[bt]
\centering
\includegraphics[width = \columnwidth]{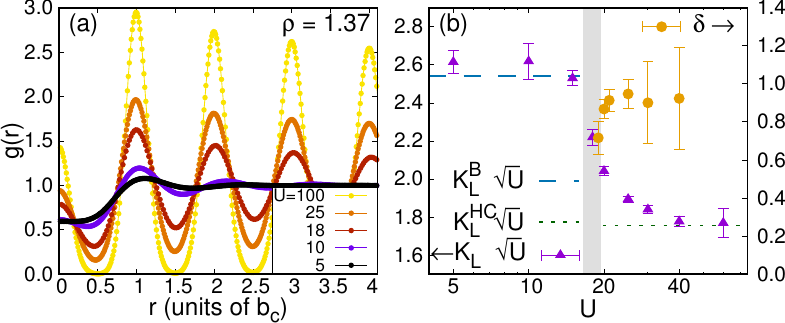}
\caption{(a) Pair distribution function at $\rho=1.37$ from the LL to the CLL phases. (b) Left y-axis: Luttinger parameter (scaled by $\sqrt{U}$) in linear-log scale (triangles), compared to Bogoliubov $K_L^B=\sqrt{3\pi\rho/2 U}$, and harmonic chain $K_L^{HC}=\sqrt{\pi^2\rho/b_c^3{\gamma}(b_c)}$ predictions. Right y-axis: excess particles per cluster $\delta$ (circles). A band highlights the transition region.}
\label{fig:gdirCL} 
\end{figure}

\begin{figure*}[ptb]
\begin{center}
\includegraphics[width =\textwidth]{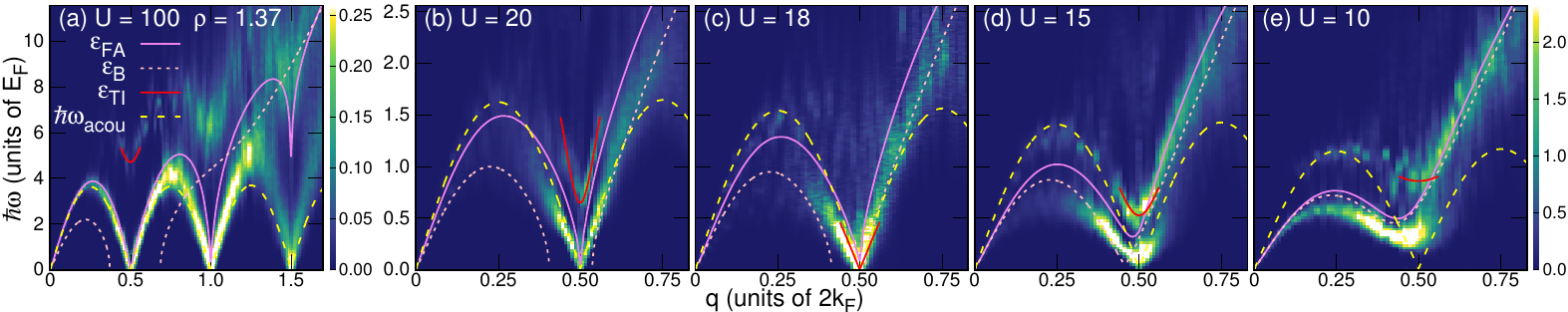}
\caption{Spectra at $\rho=1.37$ with decreasing $U$, compared to Feynman $\varepsilon_{FA}$ and Bogoliubov $\varepsilon_B$ approximations, and the harmonic chain acoustic mode ${\hbar\omega_{acou}}$. At $q\simeq q_c$, the secondary mode is fitted by the transverse Ising spectrum $\varepsilon_{TI}$, Eq.~\eqref{eq:ising}. 
Color scale as in Fig.~\ref{fig:spectraliquid}a.}
\label{fig:spectratransition} 
\end{center}
\end{figure*}

\paragraph{Commensurate Cluster Luttinger Liquid--} 
We therefore now focus on the density $\rho=2/b_c\simeq 1.37$ (solid line in Fig.~\ref{fig:phase}), commensurate to clusters of $N_c=2$ particles \cite{[{Conversely, density $\rho=1/b_c\simeq 0.68$ manifests a more standard crossover from $K_L>1$ to $K_L<1$, not shown here, but see }] [{}] otterbach_wigner_2013}. In Fig.~\ref{fig:gdirCL}a, the PIGS results for $g(r)$ are shown, indicating an evolution to a cluster structure on increasing $U$, with peaks containing two particles. $g(r)$ manifests long-range algebraic decay of the peaks' heights, 
which demonstrates absence of true crystal order \cite{mermin_absence_1966}. To interpret these results, we employ CLL theory. In the standard bosonization approach \cite{haldane_effective_1981}, the counting field fluctuates around a lattice with spacing $\rho^{-1}$: hamiltonian $H_{LL}$ is then derived with the assumption that fluctuations are small. However, in a commensurate cluster liquid, clearly fluctuations are small only around a lattice of clusters, with spacing $\rho^{-1}N_c=b_c$. We follow \cite{mattioli_cluster_2013,dalmonte_cluster_2015}, and obtain the following commensurate CLL form of $g(r)$:
\begin{equation}
 g(r)\underset{r\gg 1/\rho}{\simeq} 1-\frac{2K_L}{(2\pi\rho r)^2}+\sum_{l=1}^\infty A_l \frac{\cos{(2\pi l \rho r/N_c)}}{r^{2K_L^\prime l^2}}
\end{equation}
The $r^{-2}$ term is analog to the standard LL case, 
while the last term yields dominant density oscillations of wavevector $2k_F/N_c=q_c$, modulated by an effective Luttinger parameter $K_L^\prime=K_L/N_c^2$. This implies that, in the CLL phase, the divergence of $S(q_c)\propto N^{1-2K_L^\prime}$ is much stronger than what would result from $K_L$. We extract $K_L$ and $K_L^\prime$ from the small momentum behavior of $S(q)$ and large distance decay of $g(r)$, respectively \cite{Note1}. Interestingly, $K_L$ scales as $U^{-1/2}$ in both the LL and CLL regimes, but with different prefactors. Moreover, we verify that the number of excess particles per cluster $\delta=\sqrt{K_L/K_L^\prime}-1$ quickly goes to $1$ for $U>18$ (Fig.~\ref{fig:gdirCL}b).

Deeply in the cluster phase, a composite harmonic chain (HC) theory can also be envisaged. We write a model hamiltonian of the type ${{H}_{HC}}=\sum_{i\nu} p^2_{i,\nu}/2+\gamma\sum_{i\nu\mu}(x_{i,\nu}-x_{i+1,\mu})^2/2$, where $x_{i,\nu}$ is the displacement of the $\nu$-th particle (with $1\le\nu \le N_c$) from the average position of cluster $i$ (with $1\le i\le N/N_c$), and springs of strength $\gamma$ are present only between particles in adjacent clusters, modeling the fact that $V(r)$ is flat at short distances \footnote{PBC are implied}. We obtain center-of-mass modes, of acoustic frequencies $\omega_{acou}(k)=2\sqrt{N_c{\gamma}}\sin{(k b_c/2)}$, and optical modes, of dispersionless frequency ${\omega_{opt}}=\sqrt{2N_c{\gamma}}$. The latter correspond to relative vibrations of particles in a cluster.
We relate ${\gamma}$ to the mean-field potential felt by a particle if all the others are in a cluster crystal with spacing $b$, and find $\gamma(b)=-(4\pi^2/b^3) \sum_{j=1}^\infty j^2\tilde{V}(2\pi j/b)$ \cite{[{Remarkably, discrepancies of this simple expression from the calculation of the normal mode frequencies using the full dynamical matrix, as calculated in }] [{ are very small.}] neuhaus_phonon_2011}. It is clear that a stable structure is possible only if  $\tilde{V}(2\pi/b)<0$ for some $b$ \cite{mladek_formation_2006,likos_why_2007}. 

In Fig.~\ref{fig:spectratransition}, the spectra at $\rho=1.37$, with decreasing $U$, are shown. Panel a ($U=100$) is deep in the CLL phase: the main peak is in good agreement with the acoustic mode of HC theory, with $b=b_c$. A secondary structure appears at higher frequencies, which we interpret as the optical mode \cite{[{As also suggested in the 2D case in }] [{}] saccani_excitation_2012}. This is however not flat, but strongly modified by anharmonic couplings: these are clearly even more crucial at smaller $U$, where they induce cluster melting.

\begin{figure}[pb]
\centering
\includegraphics[width = \columnwidth]{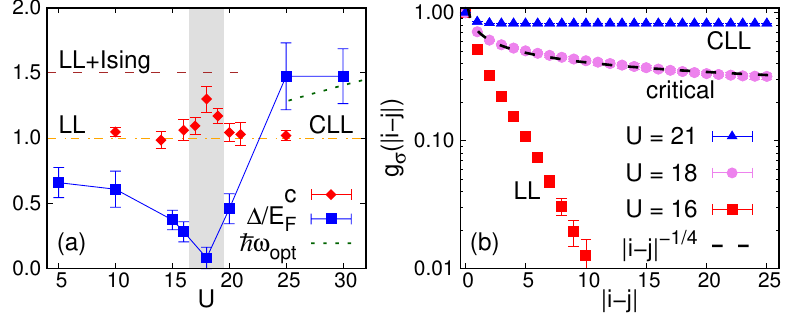}
\caption{(a) Second mode gap $\Delta$ (squares), compared to the optical mode energy ${\hbar\omega_{opt}}$ (dotted line); central charge $c$ (diamonds). (b) Log-linear scale: pseudo-spin correlators, compared to the critical behavior (dashed line).}
\label{fig:ising} 
\end{figure}

\paragraph{Ising transition--} The question is now: how are the LL and CLL phases really different? Is the transition simply a crossover? 
Our data, which show Luttinger liquid behavior on both sides, exclude Berezinskii-Kosterlitz-Thouless transition to a charge-density-wave, or Peierls transition, even though, at $U=18$, we get $K_L\simeq 0.52(1)$ \cite{giamarchi_quantum_2003}. Moreover, the atomic-pair superfluid transition \cite{romans_quantum_2004} is also excluded, since, here, formation of larger clusters is allowed
and the CLL phase manifests strong quasi-solid order ($K_L<1/2$).  

In fact, the physics of this relatively simple system is very rich. 
The acoustic mode of the CLL phase (Fig.~\ref{fig:spectratransition}, panels a-b) is gapless at $q=q_c$, corresponding to $k_F$, at this density. After the transition, to be located at $U=U_c\simeq 18$ (panel c), this lowest excitation turns into the rotonic mode (panels d-e).
Quite interestingly, a weaker secondary mode appears not only in the cluster phase, but also in the strongly correlated liquid phase, in the form of a \emph{secondary roton}, which connects to the higher-momenta main mode.
It is reasonable to associate this secondary excitation, in the LL phase, to incipient cluster formation, due to particles being preferentially localized close to either the left or the right neighbor. The crucial observation is that the gap of both such LL excitations (panels d-e), and the anharmonic optical modes of the CLL phase (a-b), vanishes at the transition (c), which implies that they proliferate at that point. 
This behavior is consistent with that of the 1D transverse Ising (TI) model \cite{pfeuty_onedimensional_1970,coldea_quantum_2010,shimshoni_quantum_2011a,ruhman_nonlocal_2012,dutta_quantum_2015} of a chain of coupled two-level systems. Its hamiltonian is $H_{TI}=-J\sum_i \sigma^z_i\sigma^z_{i+1}-h\sum_i\sigma^x_i$, where $\sigma^{x/z}_i$ are Pauli matrices at site $i$. It contains both a ferromagnetic coupling ($J>0$), which forces alignment, and quantum tunneling ($h>0$) between the eigenstates of $\sigma^z$, favoring a paramagnetic state.
This model is exactly solvable with a Jordan-Wigner transformation and Bogoliubov diagonalization \cite{lieb_two_1961} and yields excitations of energy 
\begin{equation}\label{eq:ising}
\varepsilon_{TI}(q)=\sqrt{\Delta^2+4Jh\left(\sin{{qa}/{2}}\right)^2}\;,
\end{equation}
where $\Delta=|J-h|$ is the gap and $a$ is lattice spacing, which are gapless only for $h=J$. This signals a QPT from the ferromagnetic to the paramagnetic state, which is dual to the 2D classical thermal Ising transition. In our case, it is natural to associate $\Delta$ to the gap of the secondary mode at $q=q_c$, and set $a=b_c$, implying that a spin should be identified every two particles. We fit Eq.~\eqref{eq:ising} from our spectra (Fig.~\ref{fig:ising}a): within our accuracy, the behavior of $\Delta$ in $U-U_c$ is linear close to the transition, consistent with the dynamical exponent $z=1$ \cite{dutta_quantum_2015}. The point at $U=18$ requires very long projection times: another indication of the presence of a very low-energy mode. Within our resolution, the Luttinger and critical Ising modes have the same velocity at $U=U_c$, which would imply low-energy supersymmetry \cite{huijse_emergent_2015}. 

To corroborate our interpretation, we recall that the central charge $c$ of the critical TI model is $c=1/2$, so that, at the transition, the total central charge should be $c=1+1/2=3/2$, as calculated for the related lattice model \cite{dalmonte_cluster_2015}. We estimate $c$ from the slope of the energy per particle $\varepsilon(N)=\varepsilon_\infty - c E_F/(6 K_L N^2)$ versus $1/N^2$, employing a standard CFT result for the dominant finite-size effects \cite{affleck_universal_1986,blote_conformal_1986}. An increase of $c$ is manifest in Fig.~\ref{fig:ising}a. It is in fact delicate to extrapolate $c$ close to $U_c$: higher order corrections may become relevant, and field theoretical methods should elucidate the interplay between the Luttinger and Ising fields, as done in \cite{huijse_emergent_2015,alberton_fate_2017}. It would be interesting to extract $c$ also from the entanglement entropy, as recently introduced in PIGS \cite{Herdman_Spatialentanglemententropy_2016}.

It is particularly appealing to investigate the microscopic realization of this effective TI model. The many-body potential surface reduces to double wells as a function of relative distances: two nearby bosons have preferred configurations if they are at $r_{ij}=0$ or $r_{ij}\simeq b_c$ \cite{Note1}. Thus, for each even particle, for example, there are left $\psi_L$ and right $\psi_R$ preferred cluster configurations, and the potential energy is minimized when subsequent even particles choose the same clustering direction. Anharmonic terms give instead rise to delocalization. The cluster phase is then to be thought as the ferromagnetic state, where all even particles have chosen either $\psi_L$ or $\psi_R$ (Fig.~\ref{fig:phase}b), while the liquid phase is made of $\psi_+=(\psi_L+\psi_R)/\sqrt{2}$ states (Fig.~\ref{fig:phase}a), where particles continuously hop left and right. 

This mapping can be made quantitative, by introducing a simple, but effective \emph{string} representation of $\sigma^z$, inspired by \cite{ruhman_nonlocal_2012}: 
first, particles are \emph{ordered by their position} $k$, and even positions are assigned a lattice index $i=k/2$; then, a pseudo-spin $\sigma_i^z=1$ is assigned if $|x_k-x_{k-1}|<|x_k-x_{k+1}|$, or $\sigma_i^z=-1$ in the opposite case. We evaluate the spatial correlator $g_{\sigma}(|i-j|)=\langle \sigma^z_i\sigma^z_j\rangle$ of such $\mathbb{Z}_2$ pseudo-spin (Fig.~\ref{fig:ising}b). It is very remarkable that $g_{\sigma}$ behaves as expected for the TI model: in the LL (paramagnetic) phase it decays exponentially, while in the CLL (ferromagnetic) phase it manifests true long-range order, which is \emph{nonlocal} \cite{haldane_nonlinear_1983,dallatorre_hidden_2006,ruhman_nonlocal_2012}, because of the preliminary ordering of particles. At $U=18$, its behavior is close to an algebraic decay with exponent $\eta\simeq-1/4$.

This pseudo-spin mapping in a continuous system at the LL-CLL transition, as revealed by excitation spectra and a suitable spin correlator, is the key result of this Letter. Such critical regime could be probed even at finite $T$, given finite experimental sizes. Future work will investigate effects of non commensurability of $\rho$ with $1/b_c$, which is particularly relevant for trapped gases. Also, an open issue is the presence of quantum Potts transitions at densities commensurate to $N_C\ge3$.

\begin{acknowledgments}
We acknowledge useful discussions with M. Dalmonte, M. Fleischhauer, C. Gross, R. Martinazzo, A. Parola, N. Prokof'ev, H. Weimer.
We acknowledge the CINECA awards IscraC-SOFTDYN (2015) and IscraC-CLUDYN (2017) for the availability of high performance computing resources and support. We acknowledge funding from the University of Milan (grant PSR$2015$-$1716$LPERI-M).
\end{acknowledgments}

%


\newcommand{\bb}[1]{\boldsymbol{#1}}
\newcommand{\gexp}[1]{\ensuremath{\exp\left\{#1\right\}}}
\newcommand{\mtext}[1]{\qquad \text{#1} \qquad}

\pagebreak
\onecolumngrid
\vspace{\columnsep}
\newpage
\begin{center}
\textbf{\large Supplemental Material: Quantum Critical Behavior of One-Dimensional Soft Bosons in the Continuum}
\end{center}
\vspace{2cm}
\twocolumngrid

\setcounter{equation}{0}
\setcounter{figure}{0}
\setcounter{table}{0}
\renewcommand{\theequation}{S\arabic{equation}}
\renewcommand{\thefigure}{S\arabic{figure}}
\renewcommand{\thetable}{S\arabic{table}}
\renewcommand{\bibnumfmt}[1]{[S#1]}
\renewcommand{\citenumfont}[1]{S#1}
\addtolength{\textfloatsep}{1mm}

We give details on the potential and relevant mean-field potentials; make some comments on the experimental realization; briefly explain the PIGS and GIFT algorithms, the optimization of the trial wavefunction, the extraction of the Luttinger parameters and of the central charge, providing also tables with the results used to produce Figs. 3 and 5 in the main text.  

Note: citations in this Supplemental Material refer to the bibliography in the main paper.

\section{Potential and effective mean-field potential}

\begin{figure}[btph]
\raggedright
\hspace{0.3cm}
\includegraphics[width=0.8\columnwidth]{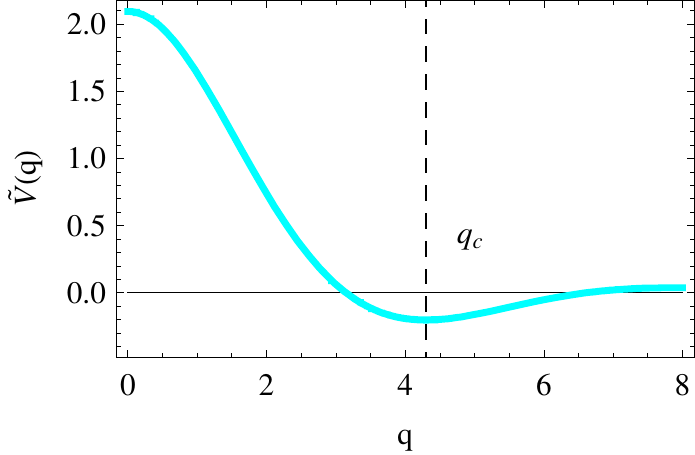}
\caption{1D Fourier transform of the shoulder potential (with $U=1$).}
\label{fig:vtilde}
\end{figure}

The soft-shoulder potential $V(r)=U/(1+r^6)$ has the 1D Fourier transform (see Fig.~\ref{fig:vtilde})
\begin{multline}\tilde{V}(q)=\int_{-\infty}^{+\infty}e^{i q r} V(r)dr=\\
{U \pi} e^{-{q/2}}\left[e^{-{q/2}} + \cos\left(\sqrt{3} q/2\right) + \sqrt{3} \sin\left(\sqrt{3} q/2\right)\right]/3\; .
\end{multline}

\begin{figure}[ptb]
	\centering
	\includegraphics[width=0.9\columnwidth]{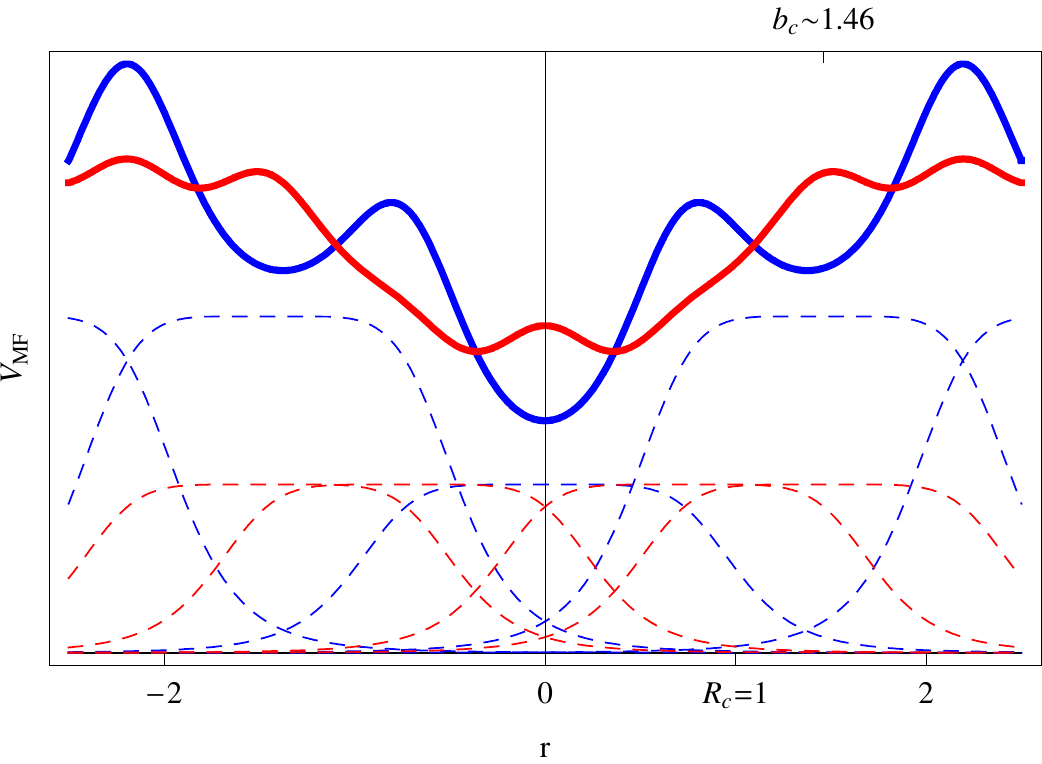}
	\caption{Approximate mean field potentials felt by a particle in the LL phase (red solid line) and in the dimer CLL phase (blue solid line). Dashed lines indicate some of the contributions to the mean-field potential (arb. energy units).}
	\label{fig:vmf}
\end{figure}

When the density is $\rho=2/b_c\simeq 1.37$, an approximation of the mean-field potential felt by a particle in the liquid phase can be obtained by letting all other particles on a lattice of spacing $1/\rho=b_c/2$: $V_{MF}^{LL}(r)=-V(r)+\sum_{i=-\infty}^{\infty} V(r - i b_c/2)$ (red solid line in Fig.~\ref{fig:vmf}). This lattice periodicity is clearly unstable, since a double well appears close to the origin: so the liquid phase is stable only thanks to kinetic energy. The mean-field potential felt by a particle in the cluster liquid phase can be approximated by letting all other particles on a lattice of spacing $2/\rho=b_c$: $V_{MF}^{CLL}(r)=-V(r)+2\sum_{i=-\infty}^{\infty} V(r - i b_c)$, which implies that pairs of particles overlap (blue solid line in Fig.~\ref{fig:vmf}). Here, a clear harmonic confinement is present at $r=0$, which stabilizes the cluster phase, and whose typical energy $\varepsilon_h(b_c)$ is given in the main text.

\begin{figure}[pbth]
	\centering
	\includegraphics[width=0.85\columnwidth]{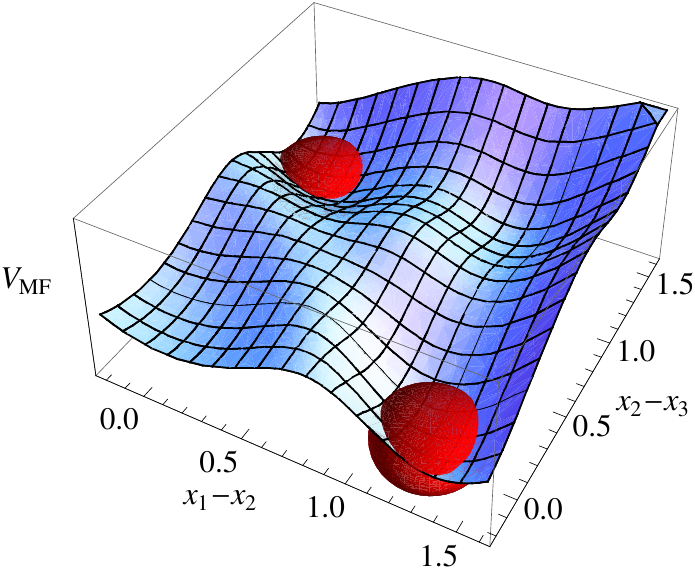}
	\caption{Approximate mean field potential felt by three particles in the LL phase. The spheres indicate the two main minima.}
	\label{fig:doublewell}
\end{figure}

An instructive mean-field potential can also be visualized, in the LL phase, by letting three particles free, and fixing only their center of mass at the center of a lattice of the other particles with spacing $1/\rho=b_c/2$: the resulting surface, in terms of the relative distances $r_{12}=|x_1-x_2|$ and $r_{23}=|x_2-x_3|$, is plotted in Fig.~\ref{fig:doublewell}. The two spheres indicate the positions of the two main equivalent minima, which clearly correspond to $(r_{12},r_{23})\simeq(0,b_c)$ and $(r_{12},r_{23})\simeq(b_c,0)$. For particle 2, suitable orbitals centered in such minima correspond to the two states $\psi_L$ and $\psi_R$ introduced in the text. They define the Hilbert space governed by the Ising hamiltonian introduced in the main text.

\section{Considerations about the experimental realization}
In the context of ultracold Rydberg gases, the parameters of the 3D shoulder potential can be related to the Rabi frequency $\Omega_R$ and detuning $\Delta_R$, by $U=\left(\frac{C_6 m^3 |\Delta_R|^2}{16\hbar^4}\right)^{1/3} \beta^4$ and $R_c=\left(\frac{C_6}{2\hbar|\Delta_R|}\right)^{1/6}$, where $\beta=\frac{\Omega_R}{2\Delta_R}$ is the admixture of the Rydberg level with the ground-state, and $C_6$ is van-der-Waals coefficient [11].
Eq.(1), in the main text, is the effective 1D Hamiltonian [21],
which is relevant in an elongated quasi-1D configuration, once the transverse degrees of freedom are frozen in the ground state, due to strong trapping [5].
The transverse confinement size $a_\perp$ does not affect much the effective potential, if $a_\perp\lesssim R_c$. With respect to the original 3D potential, the quasi-1D interaction has reduced $U$ and larger $q_c$. The condition $a_\perp\lesssim R_c$ is already under reach in current experiments, where both $a_\perp$ and $R_c$ can be of order of $100$ nm [6].
The above condition also prevents the zig-zag instability [53]
for strong interactions. To experimentally probe the phase diagram in Fig.1 of the main text, one can vary detuning $\Delta_R$, linear density $n$, mass $m$ (by changing the atomic species), and $C_6\propto \bar{n}^{11}$, by changing the addressed Rydberg level of principal number $\bar{n}$. Note that, while Rydberg levels have typical lifetimes $\tau_R$ of order of $\mu s$, Rydberg dressed atoms with small admixture $\beta$ have proportionally increased lifetimes $\tau=\tau_R/\beta^2$ [16,17].
Cryogenic techniques would be beneficial to reduce the detrimental black-body radiation, due to the experimental apparatus [8].
The experimental realization of hamiltonian of Eq.(1) in the main text is presently challenging in the absence of a lattice, since no short-range hard core is enforced. However, clustering effects persist even if one introduces a hard core, provided it is small when compared to $R_c$.

\section{Path-Integral Ground State method}

(For completeness, we add a brief description, adapted from the Supplemental Material of Ref.[38].)

The Path Integral Ground State (PIGS) Monte Carlo method is a projector 
technique that provides direct access to ground-state expectation values
of bosonic systems, given the microscopic
Hamiltonian $\hat H$ [34,35].
The method is exact, within unavoidable statistical error bars, which can nevertheless be reduced by performing longer simulations, as in all Monte Carlo methods.  Observables $\hat{O}$ are calculated as $\langle \hat{O}\rangle = \lim_{\tau\to\infty}\langle\Psi_\tau|\hat{O}|\Psi_\tau\rangle/\langle\Psi_\tau|\Psi_\tau\rangle$, 
where $\Psi_\tau = e^{-\tau \hat{H}} \Psi_T$ is the imaginary-time projection of an initial trial wave-function $\Psi_T$. Provided non-orthogonality to the ground state, the quality of the wave-function only influences the projection time practically involved in the limit and the variance of the results.

\begin{figure}[btp]
\centering
\includegraphics[width=0.85\columnwidth]{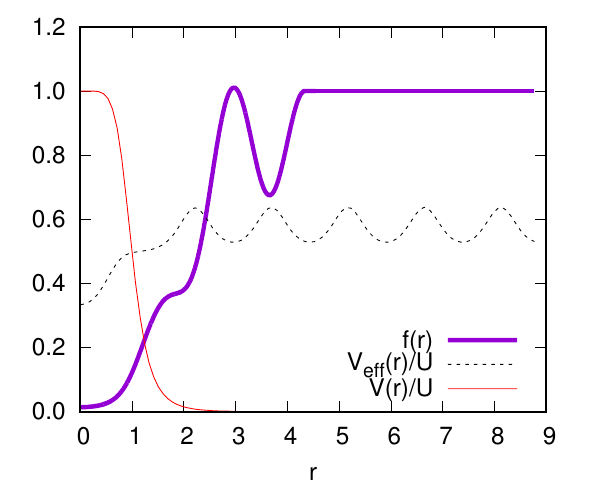}
\caption{Effective potential (dashed line), compared to the bare potential (solid thin line) and optimized Jastrow two-body correlation (solid thick line) at $U=80$.}
\label{fig:jastrow}
\end{figure}

The trial wavefunction is of the two-body Jastrow form: $\Psi_T(x_1\dots x_N) = \exp\left[-\frac{1}{2}\sum_{i<j} \left(u(r_{ij}) + \chi(r_{ij})\right)\right]$. The Reatto-Chester contribution 
\begin{equation}
\chi(r)=- \alpha \log{\left[\frac{ \sin^2{\left(\frac{\pi r}{L}\right)}+\sinh^2{\left(\frac{\pi \bar{R}}{L}\right)} }{ 1+\sinh^2{\left(\frac{\pi \bar{R}}{L}\right)} }\right]} 
\end{equation}
accounts for long-wavelength phonons [36,30]
and allows for faster convergence of the small momenta parts of the structure factors. $\bar{R}$ is a variational parameter delimiting the long-range part from the short-range contribution. The latter is embodied in the $f(r)=\exp\left[-u(r)/2\right]$ factor, which we take as the numerical solution (described in detail in [30])
of a suitable two-body Schr\"odinger equation (with reduced mass $\mu=m/2$), with the effective potential 
\begin{equation}
V_{eff}(r)=c_1 V(r) + c_2 \sum_{l=-l_C}^{l=l_C} V(r-l b) \;,\label{eq:veff}
\end{equation}
and the boundary condition $u(\bar{R})=0$. Here, the first term is a renormalized two-body potential, while the sum accounts for a series of nearby clusters in a lattice of spacing $b$ (l=0 is excluded and $l_M=int[L/(2b)]$). All parameters in the trial wave-function are variationally optimized, as described in the next Section, before projecting with the PIGS method, in order to maximize efficiency of the simulations. One finds that typically $\bar{R}\sim 2.2$,  $b\sim b_c$, and $c_1$ and $c_2$ are usually of the same order in the cluster phase. We found it is crucial to introduce such effective cluster contributions in the two-body Jastrow factor, since the variational energy per particle is typically $30\%$ higher than the PIGS result, if one sets $c_2=0$. On the contrary, by properly optimizing the parameters, we obtained discrepancies of at most a few percent of the PIGS result. A typical effective potential, and the corresponding short-range Jastrow correlation $f(r)$, are shown in Fig.~\ref{fig:jastrow}. Notice the behavior of $f(r)$ at $r\simeq i b_c$, with $i=1,2$. 

In the PIGS method, the imaginary time $\tau$ of propagation is split into $M_P$ time steps of size $\delta\tau=\tau/M_P$ so that a suitable short-time approximation for the propagator can be used; in this case, we employ the fourth-order pair-Suzuki approximation [35].

The timestep is selected by performing various simulations at short fixed projection time $\tau$ with increasing number of beads $M_P$, and monitoring the energy per particle, looking for convergence within errorbars of order $10^{-4}$. At $\rho=1.37$, the resulting $d\tau$ approximatively follows the form $d\tau\simeq 0.1 U^{-1/2}$: this is consistent with the consideration that the imaginary-time density matrix typically varies at the scale of the amplitude $l_{HCO} \propto U^{-1/4}$ of the cluster vibrational modes.

Once the timestep is selected, convergence to the ground-state is obtained by selecting a sufficiently long projection time $\tau$, which renders the potential energy flat (within errorbars) as a function of time slice. With our optimized initial wavefunctions, the typical number of time-steps for convergence is $35$.
Once convergence in $\tau$ is obtained, a further projection time of typically $M_F=60$ time-slices is used to sample the intermediate scattering function $F(q,\tau)$.

Close to the transition, convergence in $\tau$ is very slow, due to the low-energy Ising mode. We have thus also employed a more refined effective potential, of the form
\begin{multline}
V_{eff}(r)=c_1 V(r)  \\
+c_2 \sum_{l=-l_C}^{l=l_C} \sum_{i=-p}^{p} V(r-l b-i\delta_l) g(i\delta_l,s_l) \;,\label{eq:veff2}
\end{multline}
where we introduced a gaussian kernel $g(r,s_l)=\exp{[-r^2/(2s_l^2)]}/\mathcal{N}(s_l)$, with $\mathcal{N}(s_l)=\sum_{i=-p}^{p} \exp{[-(i\delta_l)^2/(2s_l^2)]}$, for better describing the clusters. The width $s_l=s_0 l^{\gamma}$ is allowed to increase with the cluster position, and the convolution is discretized with $p=10$ and $\delta_l=3s_l/p$. In the limit $s_0\to0$ and $\gamma\to0$, one recovers expression \eqref{eq:veff}, while in general $s_0$ and $\gamma$ are two additional variational parameters to be optimized.

For $U=18$, we employed the parameters $c_1=1.127$, $c_2=1.076$, $b=1.655$, $s_0=0.2$, $\alpha=4.84$, $\bar{R}=1.83$, and $\gamma=0.4$. With $d\tau=0.03$, simulations still required $100$ time-slices of projection.

\begin{figure}[tbp]
\centering
\includegraphics[width=0.85\columnwidth]{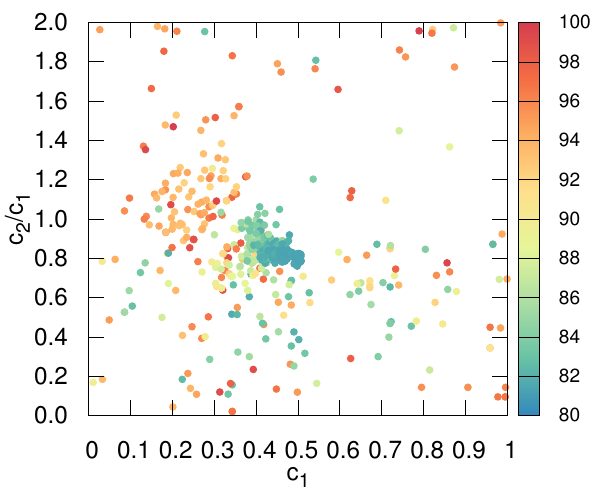}
\caption{Convergence of the two parameters $ c_1 $ and $ c_2 $ in a simulated annealing procedure at $U=100$, $\rho=1.37$. The color scale refers to the energy value of each VMC simulation.}
\label{fig:sma}
\end{figure}

\section{Parameters optimization through simulated annealing}

The parameters appearing in the trial wavefunction are optimized with the Variational Monte Carlo method (VMC), which corresponds to PIGS, when no imaginary-time projection is made. We employ a simulated annealing procedure in parameters' space, aiming at both reducing the energy associated to the trial wavefunction, and obtaining a good pair distribution function $ g(r) $. To each configuration $ \bb{\xi} $ of the parameters, the following Boltzmann weight $\lambda =  \gexp{-\beta \left[E\left(\bb{\xi}\right) + \zeta \,\chi^2\left(\bb{\xi}\right)\right]}$ is associated, with inverse temperature $ \beta $, where $E(\bb{\xi})$ is the VMC energy, and
\begin{equation}
 \chi^2(\bb{\xi}) = \sum_{i}^{i_{M}} \frac{\left[g_{0}\left(r_i\right) - g_{\xi}\left(r_i\right)\right]^2}{\sigma_{0}^2\left(r_i\right) + \sigma_{\xi}^2\left(r_i\right)}
\end{equation}
is the discrepancy between the pair distribution functions obtained at a particular step $ g_{\xi} $, and the one obtained with a preliminary unoptimized PIGS simulation $ g_0 $, weighted by the corresponding uncertainties $ \sigma_{\xi} $ and $ \sigma_{0} $. This difference is evaluated only up to the $i_{M}$-th bin of the histogram of $g(r)$, which includes the first peak of $g(r)$ for $r\neq 0$. The motion of a fictitious particle is simulated in parameters' space, subjected to such Boltzmann weight, where $\zeta$ is the relative role of $\chi^2$ with respect to $E$. The effective temperature is progressively decreased in order to reach the configuration having the lowest value possible of $ E(\bb{\xi}) $ and $ \chi(\bb{\xi}) $ simultaneously, and the algorithm is stopped once the variations of the weight are of the order of errorbars. The typical used value of $ \zeta $ is $ \sim 10^{-4} $, except close to the transition, where we observed that increasing this value led to shorter projection times in the subsequent PIGS simulations. Fig.~\ref{fig:sma} shows an example of the convergence of the two parameters $ c_1 $ and $ c_2 $.

\section{Genetic Inversion via Falsification of Theories method}

(For completeness, we add a brief description, adapted from the Supplemental Material of Ref.[38].)

The relation between the intermediate scattering function $F$, which is evaluated within the PIGS simulations, and the dynamical structure function is
\begin{equation}
\label{eq:laplace}
F(q,\tau) = \frac{1}{N} \langle e^{\tau H} \rho_q e^{-\tau H} \rho_{-q} \rangle = \int_0^\infty d\omega e^{-\tau\omega} S(q,\omega).
\end{equation}

This is a Fredholm equation of the first kind and is an ill-conditioned problem, because a small variation in the imaginary-time intermediate scattering function $F$ produces a large variation in the dynamical structure factor $S$. At fixed momentum $q$, the computed values $F_j=F(q,j\delta\tau)$, where $j=0\dots M$, are inherently affected by statistical uncertainties $\delta F_j$, which hinder the possibility of deterministically infer a single $S(q,\omega)$, without any other assumption on the solution. The Genetic Inversion via Falsification of Theories method (GIFT) exploits the information contained in the uncertainties to randomly generate $Q$ compatible instances of the scattering function $F^{(z)}$, with $z=1\dots Q$, which are independently analyzed to infer $Q$ corresponding spectra $S^{(z)}$, whose \emph{average} is taken to be the ``solution''. This averaging procedure, which typifies the class of stochastic search methods (See the recent review [40]),
yields more accurate estimates of the spectral function than standard Maximum Entropy techniques. This method is able to resolve the first spectral peak and approximately the second one, if the errorbars of $F$ are sufficiently small; generally speaking, the most relevant features of the spectrum are retrieved in their position and (integrated) weight. 

Given an instance $z$, the procedure of analytic continuation from $F^{(z)}$ to $S^{(z)}$ relies on a stochastic genetic evolution of a population of spectral functions of the generic type $S^{(z)}(q,\omega) = m_0\sum_{i=1}^{N_\omega}s_i\delta(\omega-\omega_i)$, where $m_0=F^{(z)}(q,0)$ and the zeroth momentum sum rule $\sum_{i=1}^{N_\omega}s_i=1$ holds. The $N_\omega$ support frequencies $\omega_i$ are spaced linearly up to an intermediate frequency $\omega_m$, which is of order of the maximum considered momentum squared $q^2$, then a logarithmic scale is used up to very high frequencies. Genetic algorithms provide an extremely efficient tool to explore a sample space by a non-local stochastic dynamics, via a survival-to-compatibility evolutionary process
mimicking the natural selection rules; such evolution aims toward increasing the \emph{fitness} of the individuals, defined as
\begin{multline}
\Phi^{(z)}(S) =
-\sum_{j=0}^M \frac{1}{\delta F_j^2}\left[F^{(z)}_j - m_0\sum_{i=1}^{N_\omega} e^{-j \delta\tau\omega_i}\,s_i\right]^2
\\
- \epsilon \left[m_1 - m_0\sum_{i=1}^{N_\omega} \omega_i\,s_i\right]^2 \;,
\label{fitness}
\end{multline}
where the first contribution favors adherence to the data, while the second one favors the fulfillment of the f-sum rule, with $m_1={q^2}/2$ and $\epsilon$ a parameter to be tuned for efficiency. A step in the genetic evolution replaces the population of spectral functions with a new generation, by means of the ``biological-like'' processes of {\it selection}, {\it crossover}, {\it local mutation}, {\it non-local mutation}, which are described in detail in [37,40].
Moreover, the genetic evolution is tempered by an acceptance/rejection step based on a reference distribution $p^{(z)}(S)=\exp{(\Phi^{(z)}(S)/T)}$, where the coefficient $T$ is used as an effective temperature in a standard simulated annealing procedure.
We found that this combination is optimal in that it combines the speed of the genetic algorithm with the prevention of strong mutation-biases thanks to the simulated annealing.
Convergence is reached once $|\Phi^{(z)}(S)|< 1$, and the best individual, in the sense that it does not falsify the theory represented by \eqref{fitness}, is chosen as the representative $S^{(z)}$. The final spectrum is obtained by taking the average over instances $\bar{S}(q,\omega)=\frac{1}{Q}\sum_{z=1}^Q S^{(z)}(q,\omega)$.

\begin{figure}[tbp]
	\centering
	\includegraphics[width=0.85\columnwidth]{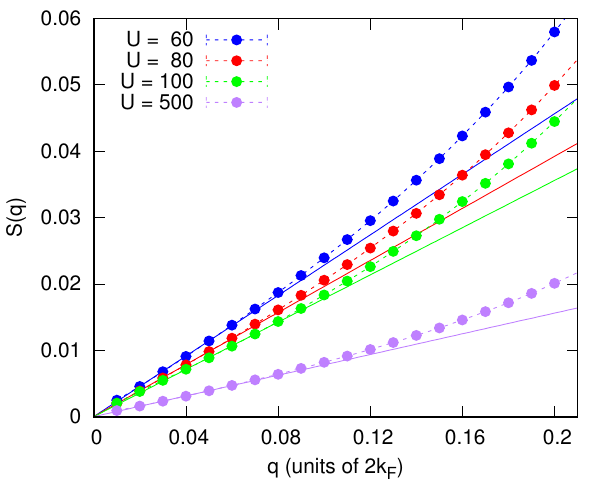}
	\caption{Low momenta behavior of $ S(q) $ and corresponding linear fit, with $\rho=1.37$.}
	\label{fig:skx}
\end{figure}

\begin{figure}[btp]
\centering
	\includegraphics[width=0.9\columnwidth]{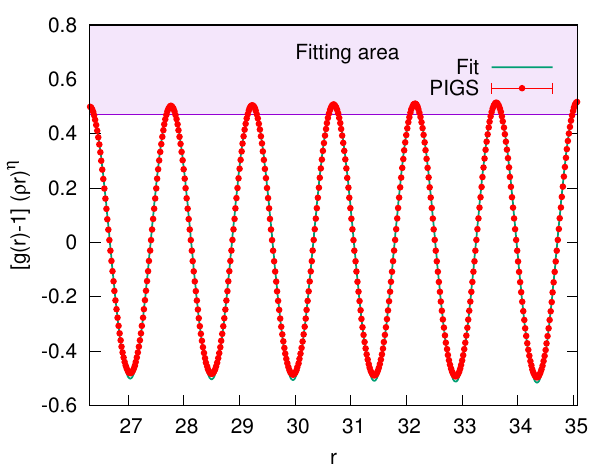}
	\caption{Fit of the decay of $ g(r) $ for $ U=25 $, $\rho=1.37$, using Eq.(2) in the main text. The purple area highlights the fitting range.}
	\label{fig:gdir}
\end{figure}

\section{Extraction of the Luttinger parameter from the static structure factor}

The behavior of $ S(q) $ for small momenta is shown in Figure \ref{fig:skx}. Since its Luttinger liquid behavior is $ S(q) \sim qK_L/(2k_F)$, we fit $ 2 k_F S(q)/q $ with a constant. Clearly, for extremely low momenta, size effects become relevant, since finite imaginary-time projection has not been able to correctly reconstruct the long-distance density fluctuations, leading to a non linear behavior of the data. For this reason, the error on the fitted value of $ K_L $ has been taken to be the absolute value of the difference between the fitted $ K_L $ and the value of $ 2k_F S(q)/q $ for the smallest $q$ at our disposal. The resulting $K_L$ are summarized in Table~\ref{tab:KL}.

\section{Extraction of the anomalous Luttinger parameter from the pair distribution function}

CLL theory implies that the long-range behavior of $g(r)$ has to be of the form of Eq.(2) in the main text. The easiest way to extract $ K_L^\prime $ from $ g(r) $ consists in evaluating the power-law exponent relative to the cosine decay. In such procedure we have neglected the $ -2K_L/(2\pi\rho r)^2 $ term, as well as all the cosine terms, but the first one, as their decay is usually too fast to be relevant. Values of $ g(r) <1$ are determined by short-range effects: it is thus appropriate to fit just the upper part of the peaks of $g(r)$. To make such procedure more precise, we have fitted the function $ \left[g(r) - 1\right] \cdot (\rho r)^{\eta} $, which is supposed to behave like $ \sim A_1 \cos\left(\pi\rho r\right) (\rho r)^{\eta -2K_L'} $ for long distances, with $ \eta $ tuned as to have the periodic peaks at comparable heights. Thus the cosine decay is considerably softened, leading to an easier fitting of $ K_L^\prime $. An example of one of our fits is shown in figure \ref{fig:gdir}. We have not been able to apply this methodology for $U<18$, due to large noise-to-signal ratio, and at $U=18$, possibly due to the role of subleading terms in the decay. The resulting $K_L^\prime$ are summarized in Table~\ref{tab:KL}.
\vspace{0.5cm}

\begin{table}[tb]
\begin{tabular}{|l|l|l|l|l|}
\hline\hline
$U$   & $K_L$  & $K_L^\prime$  & $\Delta/E_F$  & $c$     \\
\hline
5 & 1.17(3)  & -          & 0.66(11)    &   -      \\
10 & 0.83(3) & -          &   0.61(14)  &  1.05(4) \\
14 & -       & -          &  -          &  0.98(7) \\
15 & 0.65(1) & -          &  0.37(7)    &    -     \\
16 & -       & -          & 0.28(8)     &  1.06(8) \\
17 & -       & -          &   -         &  1.09(7) \\
18 & 0.52(1) & -          &  0.08(8)    &  1.3(1) \\
19 & 0.480(4) & 0.163(15) &    -        &  1.17(6) \\
20 & 0.455(5) & 0.130(6)  & 0.46(11)    &  1.04(7) \\
21 & 0.430(3) & 0.117(6)  &     -       &  1.03(9) \\
25 & 0.379(3) & 0.100(7)  & 1.47(26)    &  1.02(4) \\
30 & 0.336(4) & 0.09(2)   & 1.47(21)    &   -      \\
40 & 0.281(3) & 0.08(2)   &     -       & -        \\
60 & 0.23(1)  & -         &     -       & -       \\
\hline\hline
\end{tabular}
\caption{Values of Luttinger parameters $K_L$ and $K_L^\prime$ (Fig.3b in the main text), of the Ising gap $\Delta/E_F$ and the central charge $c$ (Fig.5a in the main text), at $\rho=1.37$.}
\label{tab:KL}
\end{table}

\section{Extraction of the central charge}

\begin{figure}[b]
	\includegraphics[width=0.95\columnwidth]{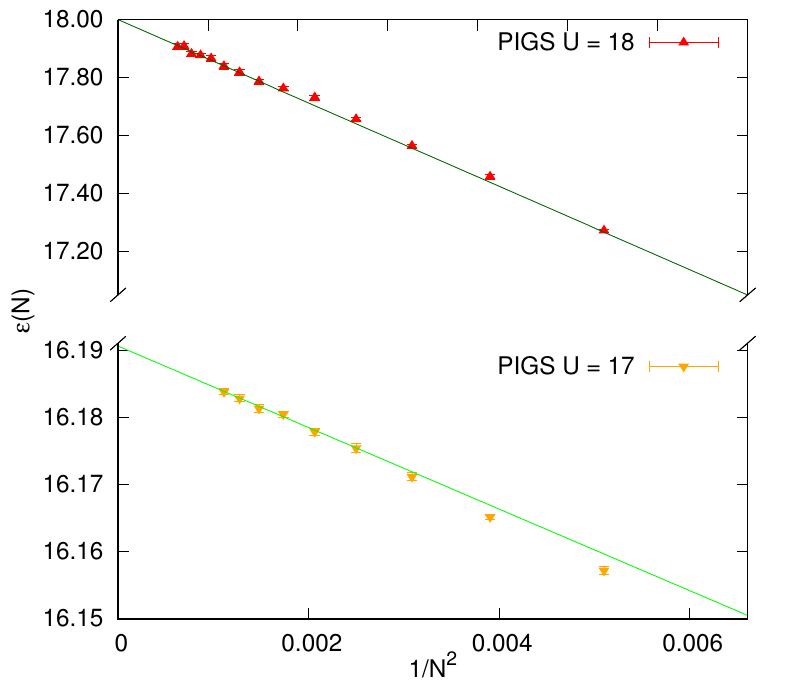}
	\caption{Example of the extraction of the central charge from the energy per particle $\varepsilon(N)$, for $U=17$ (bottom) and $U=18$ (top).}
	\label{fig:central}
\end{figure}

We estimate the central charge $c$ by fitting the slope of the energy per particle $\varepsilon(N)=\varepsilon_\infty - c E_F/(6 K_L N^2)$ versus $1/N^2$, where $K_L$ is inferred from Table~\ref{tab:KL}. The range of considered numbers of particles is typically from $N=10$ to $N=40$ ($N=100$ in some cases), and we progressively increase the minimal $N_{\text{min}}$ used in the fit, to assess the role of higher order contributions, finding that the extracted $c$ is generically stable if $N_{\text{min}}=20$. The errorbar comes from the uncertainty in $\varepsilon(N)$ and $K_L$, and from the variance of the extracted $c$ at different $N_{\text{min}}>20$.
Examples of extraction of $c$ are shown in Fig.~\ref{fig:central}, corresponding to $U=17$ and to the more difficult $U=18$ point. Table~\ref{tab:KL} summarizes the results.

\clearpage
\end{document}